\documentclass[prl,twocolumn,floatfix,amssymb,showpacs]{revtex4}

\usepackage{graphicx}

\newcommand{\ds}{\displaystyle}
\newcommand{\ud}{{\mathrm{d}}} 

\newcommand{\sd}{{\mathrm{sd}}}

\begin{document}

\title{Conductance distribution in nanometer-sized
semiconductor devices due to dopant statistics}

\author{G.~D.~J.~Smit}
\email{g.d.j.smit@tnw.tudelft.nl}
\author{S.~Rogge}
\email{s.rogge@tnw.tudelft.nl}
\author{J.~Caro}
\author{T.~M.~Klapwijk}
\affiliation{Department of NanoScience, Delft University of
Technology, Lorentzweg 1, 2628 CJ Delft, The Netherlands}

\begin{abstract}
We show that individual dopant atoms dominate the transport
characteristics of nanometer sized devices, by investigating metal
semiconductor diodes down to 15~nm diameter. Room temperature
measurements reveal a strongly increasing scatter in the
device-to-device conductance towards smaller device sizes. The
low-temperature measurements exhibit pronounced features, caused
by resonant tunneling through electronic states of individual
dopant atoms. We demonstrate by a statistical analysis that this
behavior can be explained by the presence of randomly distributed
individual dopant atoms in the space charge region.
\end{abstract}

\date{\today}


\maketitle

\section{Introduction}

In semiconductor physics, the influence of doping is generally
accounted for by a homogeneous shift of the Fermi-level, caused by
the introduction of free carriers in the semiconductor. However,
when the dimensions of a device are small compared to the average
distance between individual dopants, the discrete nature of doping
must be taken into account. Each individual ionized dopant
introduces a Coulomb potential well in the semiconductor, locally
distorting the potential landscape. When the number of dopants in
the volume of semiconductor that determines the transport
characteristics of a device gets small, these random potential
fluctuations cause atypical behavior of semiconductor devices
\cite{calvet02}. This is commonly viewed as one of the fundamental
limits in the ongoing size-reduction of CMOS-technology
\cite{meindl01,hoeneisen72}. Mapping the positions of individual
dopants \cite{voyles02} and the potential fluctuations they induce
\cite{modesti03,rau99} has been performed experimentally with
various techniques. Furthermore, the influence of statistical
fluctuations due to random dopants on device behavior has been
subject of simulations \cite{sano01}.

In this work, we experimentally investigate the effects of the
discreteness of doping on the transport properties of small diodes
by comparing many identically prepared devices. We find that
statistical fluctuations caused by randomly positioned individual
dopant atoms do not average out for very small devices. In
contrast, fluctuations dominate the electrical transport
properties of the smallest devices and cause large differences in
the conductance of nominally equal devices. Furthermore, we
demonstrate that at low temperature the Coulomb well of a single
dopant gives rise to a resonant tunneling channel.

\section{Measurements}

In order to allow for measuring the transport characteristics of
many identically prepared diodes, we use self-assembly methods to
fabricate epitaxial CoSi$_2$-diodes. The tip of a scanning
tunneling microscope (STM) is used to characterize and access the
devices individually.

All experiments are performed in an ultra-high vacuum (UHV) system
with a base pressure of $5\cdot 10^{-11}$~mbar. Self assembled
CoSi$_2$-islands are grown on Si-substrates (resistivity
$0.015~\Omega$cm, doping concentration around $2\cdot
10^{18}$~cm$^{-3}$) by evaporating a sub-monolayer of cobalt onto
a clean $7\times 7$-reconstructed Si(111) surface, followed by an
anneal at 800~$^\circ$C for about 5~min. The resulting
hexagon-shaped epitaxial CoSi$_2$-islands have heights in the
range of 2--10~nm and diameters in the range of 15--80~nm. The
inset of Fig.~\ref{fig:meanstd} displays an STM-image of a typical
island. Each island is regarded as a nanometer sized epitaxial
metal-semiconductor diode. The inter-island distances are roughly
ten times larger than the island diameters. To minimize the effect
of surface related transport channels, the $7\times 7$ surface
reconstruction surrounding the islands is destroyed by exposing it
to atomic hydrogen for 10~min at a substrate temperature of
400~$^\circ$C.

Current-voltage ($IV$) measurements are performed by positioning
the STM-tip over an island. After switching off the feedback loop,
the tip is lowered by 15~\AA, which is found to be sufficient to
make contact to the island. Then the current is measured while
ramping the voltage, yielding well-reproducible $IV$-curves that
reflect the properties of the metal-semiconductor contact. The
sample preparation and measurements techniques are described in
more detail elsewhere \cite{smit02}.

\subsection{Room temperature measurements}

\begin{figure}
  \centering
  \includegraphics[width=8.5cm]{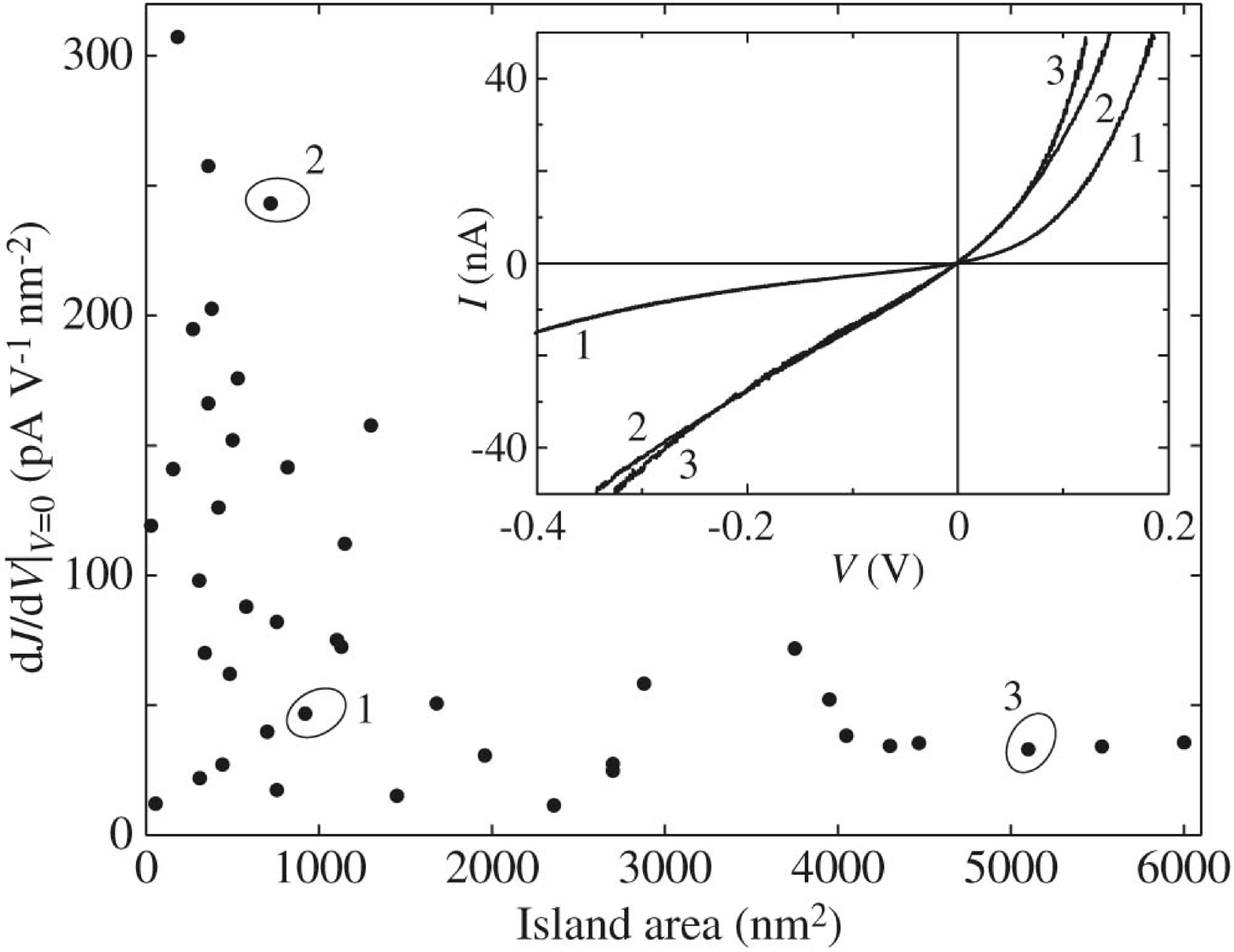}
  \caption{The zero bias conductance per unit area as a function
  of island area at room temperature. For large islands, the measured
  values fall within a narrow range. For smaller islands, the scatter
  is rapidly increasing and is much larger than the
  measurement-inaccuracy. The inset shows some typical $IV$-curves.
  The numbers in the inset correspond to the numbered data points
  in the main figure.}
  \label{fig:scatterhigh}
\end{figure}

Some typical room-temperature measurements are displayed in the
inset of Fig.\ref{fig:scatterhigh}, showing weakly rectifying
$IV$-curves, compatible with the $p$-type substrate-doping. The
overall shape of the curves is very similar. To investigate the
dependence of the diode's conductances on their size, in
Fig.~\ref{fig:scatterhigh} the zero-bias conductance per unit area
is plotted as a function of island area for more than 40 different
islands, grown on two similar $p$-type (boron-doped) samples. One
would expect that devices of equal size yield the same value for
this quantity. Indeed, for larger devices the conductance per unit
area falls within a narrow range. However, towards smaller areas
there is no definite behavior. Instead, the scatter in the
measured values increases rapidly and nominally equal devices
yield very different results.

\begin{figure}
  \centering
  \includegraphics[width=8.5cm]{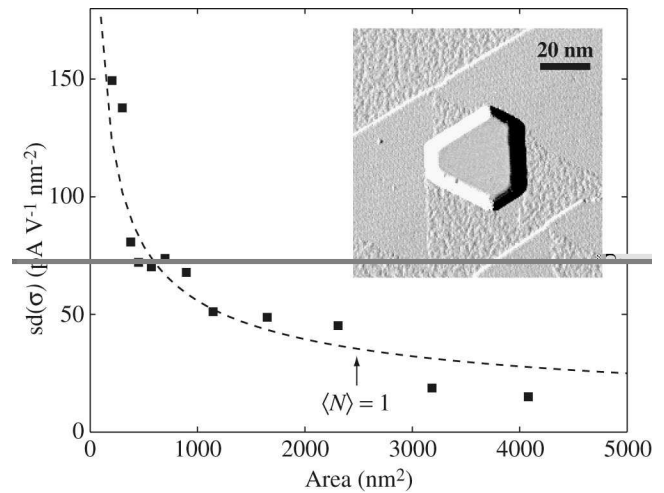}
  \caption{Standard deviations of the measured conductance per unit
  area for various values of the device area $A$. Each point is
  calculated from eight neighboring data points in
  Fig.~\ref{fig:scatterhigh}. The dashed line is a least square fit of
  the function $C/\sqrt{A}$ to these points, yielding
  $C=1.8$~nA/V/nm. The arrow indicates which island size corresponds
  to an average of one dopant atom per island ($\langle N\rangle=1$) according
  to our analysis. The inset shows an STM-image
  of a typical island, acquired at room-temperature directly after
  preparation.}
  \label{fig:meanstd}
\end{figure}

To study the increasing scatter in more detail, the standard
deviation of the conductance per unit area is plotted in
Fig.~\ref{fig:meanstd} as a function of diode area, calculated
from the data of Fig.~\ref{fig:scatterhigh}. We clearly observe
that the standard deviation increases with decreasing island size.

Measurement inaccuracies are not the source of this random
scatter, as proved by the low noise level (typically less than
10~pA around zero bias) and good reproducibility of measurements
on the same device. Moreover, this scatter is absent in similar
measurements on low-doped samples (not shown). Schottky barrier
inhomogeneities \cite{tung92}, which have been observed at
non-epitaxial interfaces \cite{talin94,meyer97,jager03}, do not
play a role here, as they are due to variations in the atomic
arrangement at the metal-semiconductor interface itself. The
metal-semiconductor interfaces in our devices are perfectly
epitaxial and mono-crystalline, which is supported by
cross-sectional transmission electron micrographs of similar
structures \cite{bennett01}. This also rules out interface defects
or grain boundaries as the origin of the fluctuations. In
addition, BEEM measurements of CoSi$_2$-films on undoped Si(111)
reveal a perfectly homogeneous Schottky barrier height
\cite{meyer97}. Only islands showing an atomically flat and
defect-free surface in the STM-images were included in the
analysis.

We will show that the increased scatter in the measured data is
caused by the presence of randomly distributed dopant atoms in the
Schottky barrier. An ionized dopant atom locally distorts the
barrier, giving rise to a local barrier reduction and thus a high
conductance spot (see the inset of Fig.~\ref{fig:LTIV}). Then, the
observed spread in Fig.~\ref{fig:meanstd} is directly related to
the spread in the number of dopant atoms $N$ in the barrier of a
device. From Poisson statistics, the relative spread is given by
$$
  \frac{\sd(N)}{\langle N\rangle}=\frac{1}{\sqrt{\langle N\rangle}},
$$
where $\sd(N)$ is the standard deviation of $N$. Because $\langle
N \rangle$ is proportional to the island area $A$, this gives that
the standard deviation of the number of dopants per unit area
increases when the device area decreases. Indeed,
Fig.~\ref{fig:meanstd} shows that a function of the form
$C/\sqrt{A}$ describes our observations appropriately.

\subsection{Low temperature measurements}

\begin{figure}
  \centering
  \includegraphics[width=8.5cm]{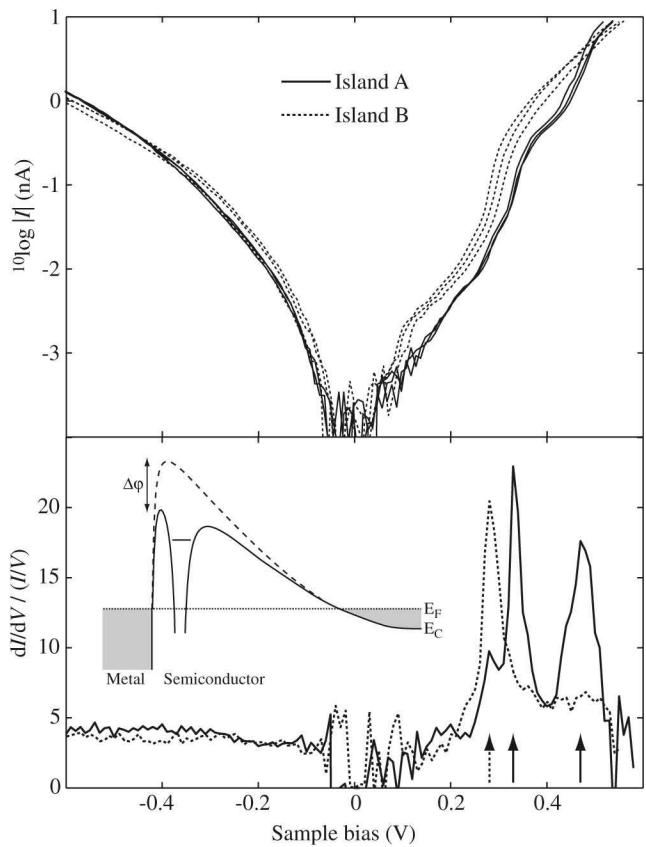}
  \caption{Measured $IV$-curves of two different CoSi$_2$-islands
  on the same sample. The upper graph shows three measurement of
  one island (island A, $A=1300$~nm$^2$) and also three measurements of
  another island (island B, $A=1500$~nm$^2$), demonstrating the
  reproducibility of the measurements. Features
  occurring at $|V|\lesssim 0.1$~V ($|I|\lesssim 1$~pA) are due to noise.
  The lower panel shows the normalized
  differential conductance of the same two islands, calculated from
  an averaged $IV$-curve. Arrows indicate the main features in the
  $IV$-curves, showing up as peaks in the lower panel.
  The inset schematically shows the band diagram of an unbiased device.
  The dashed line is the
  initial conduction band ($E_\mathrm{C}$) profile. The solid line
  is $E_\mathrm{C}$ perturbed by the dopant's Coulomb well,
  causing a local barrier lowering $\Delta\varphi$.}
  \label{fig:LTIV}
\end{figure}

To further investigate the influence of individual dopant atoms on
the transport properties of small diodes, we performed similar
experiments in a low-temperature UHV-STM operating at 4.5~K. In
the upper part of Fig.~\ref{fig:LTIV}, several $IV$-curves of two
different islands are plotted. Each represents a full and
independent measurement cycle (stop scanning, make contact, ramp
$V$, release contact, resume scanning), demonstrating their
reproducibility. The current is plotted on a logarithmic scale, to
make the features of the curves visible over several orders of
magnitude in current. Note that the noise-level is below 1\,pA.

From the measured $IV$-curves, the quantity $(V/I)\cdot(\ud I/\ud
V)$ (the so-called normalized differential conductance) has been
calculated. It was calculated from the average of several
$IV$-curves per island and plotted for the same two islands in the
lower part of Fig.~\ref{fig:LTIV}. This is a well-known approach
in the field of scanning tunnelling spectroscopy \cite{stroscio86}
and is very useful as it reduces the overall exponential behavior
and enhances bias-dependent features. Indeed, the weak features in
the $IV$-curves turn into clear peaks and the peaks show a
one-to-one correspondence to the features in the $IV$-curves. As
can be seen from Fig.~\ref{fig:LTIV}, features appear at one bias
polarity, only. This is true for all acquired curves on the same
sample. For substrates with the opposite doping type, the peaks
occur at the opposite bias polarity (not shown).

The peaks can be explained by resonant tunneling through a
discrete energy level of a dopant atom, occurring when the
Fermi-level at either side of the barrier lines up with an energy
level of the dopant's potential well. The resulting resonant
channel produces a feature in the measurement curve. Resonances
are expected at bias voltages from zero upto the barrier height
(roughly 0.5--0.7~V, depending on the dopant type). This is
consistent with the observations. The actual bias voltage at which
a resonance occurs is predominantly determined by the distance of
the dopant atom to the interface. The number of peaks in a typical
spectrum (1 to 4) corresponds to the expected number of dopants in
the devices, as we will show later. This confirms our hypothesis
that individual dopants influence the conduction path.

\section{Analysis}

Motivated by the foregoing observations, we present a model that
links the conductance fluctuations observed at room temperature to
(random) dopant positions. Moreover, the parameters in the model
are directly related to device parameters.

Because the dopants in the substrate are randomly distributed, the
number of dopants $N$ in the barrier of a given island is
Poisson-distributed with parameter $\lambda$. This means that the
probability $P_\lambda(k)$ that $N$ equals $k$ for a certain
device is given by
\begin{equation}
  P_\lambda(k)=\frac{\lambda^k}{k!}e^{-\lambda},
  \label{eq:pois}
\end{equation}
where $\lambda=\langle N\rangle$ is the mean value of the
number of dopants in the barrier of this particular device. The
parameter can be expressed as $\lambda=AtN_\ud$, where $A$ is the
area of the island, $t$ the effective thickness of the barrier and
$N_\ud$ the average doping concentration in the barrier. As shown
in Fig.~\ref{fig:meanstd}, the scatter in the number of dopants in
the barrier per unit area satisfies $\sd(N)/A\propto 1/\sqrt{A}$
and therefore is large for small islands.

We will now show that this effect also creates scatter in
conductance measurements. We assume that dopants located in a
certain region close to the metal-semiconductor interface induce a
local barrier lowering that gives rise to a low-resistance
transport channel \footnote{Note that here the term `channel'
refers to a weak spot in a barrier and is unrelated to the
conduction channels occurring in the field of quantum transport.}.
The conductance of a transport channel induced by a dopant atom
depends on the dopant's distance to the interface. Since the
distance is a random variable, the conductance $G_1$ of a single
channel is not the same for all channels, but has a certain
probability distribution. The distribution is given by a
probability density function $f_1(g)$, meaning that $f_1(g)\ud g$
is the probability for a certain channel to have a conductance
$G_1$ between $g$ and $g+\ud g$. Both the distribution of the
position of a dopant in the barrier and the dependence of the
conductance of a channel on that position are contained in
$f_1(g)$.

Assuming that the values of the conductance of the individual
channels are independent and characterized by the same
distribution given by $f_1(g)$, the total conductance of a device
is given by the sum of the contributions of the individual
channels. Here, we neglect the background conduction and assume
that the conductance is dominated by these channels. If there are
$k$ channels, the total conductance is $G_k=\sum_{i=1}^k
G_1^{(i)}$, where the conductance of the individual channels is
denoted by $G_1^{(i)}$. The density function $f_k(g)$ of $G_k$ can
be calculated explicitly by taking the $k$-fold convolution of
$f_1(g)$ \cite{grimmett92}
$$
  f_k(g)=\underbrace{f_1(g) \ast \ldots \ast f_1(g)}_{k\ \mathrm{times}}.
$$
Finally, taking into account the Poisson distribution of the
number of channels (Eq.~\ref{eq:pois}), the density function
$f(g,\lambda)$ of the total conductance $G$ of a device can be
computed as
\begin{equation}
  f(g,\lambda)=\sum_{k=0}^\infty P_\lambda(k)f_k(g).
  \label{eq:dist}
\end{equation}
In other words, for given device parameters $N_\ud$, $A$ and $t$,
$f(g,\lambda)\ud g$ is the probability that the total conductance
$G$ due to dopant atoms in the diode is between $g$ and $g+\ud g$.

\begin{figure}
  \centering
  \includegraphics[width=8.5cm]{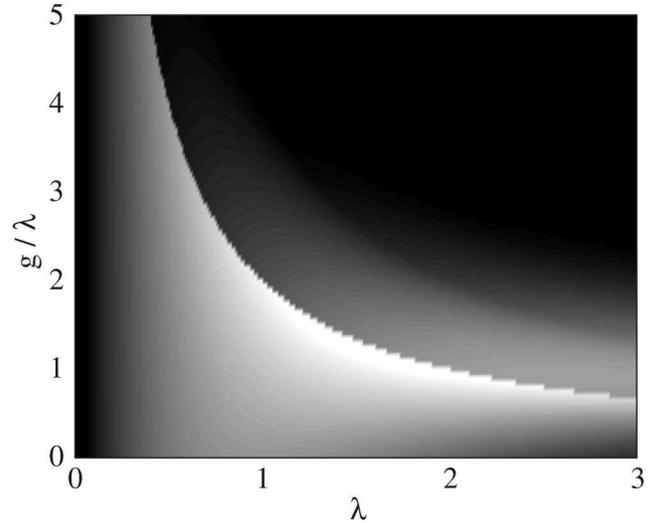}
  \caption{Map of $f(g,\lambda)$ as defined in
  Eq.~(\ref{eq:dist}) plotted as a function of
  $g/\lambda$ (vertical axis) and $\lambda$ (horizontal axis). The
  function values are represented in a linear gray-scale, where black
  corresponds to 0 and white to 0.65. For
  $f_1(g)$ a uniform distribution was chosen.}
  \label{fig:scattermap}
\end{figure}

To illustrate its behavior, Fig.~\ref{fig:scattermap} shows a map
of $f(g, \lambda)$ (from Eq.~\ref{eq:dist}) as a function of
$g/\lambda$ (proportional to the conductance per unit area) and
$\lambda$ (which is proportional to $A$). For a fixed value of the
device area (that is a fixed value of $\lambda$ and a vertical
line in the plot) the color scale gives the probability density to
find a device with a particular conductance per unit area
$g/\lambda$. For large values of $\lambda$, all density is
concentrated around $g/\lambda=\frac{1}{2}$, while for smaller
$\lambda$ it is spread over an increasingly wide range of values.
In this figure $f_1(g)$ was chosen as a uniform distribution.
However, as we will show next, the general properties of
$f(g,\lambda)$ are not strongly dependent on the particular choice
of $f_1(g)$.

Without making any assumptions on the choice of $f_1(g)$, we can
compute the moments of $f(g,\lambda)$ in terms of those of
$f_1(g)$. First, the mean value of $G$ satisfies
$$
  \begin{array}{rl}
    \langle G\rangle=&\ds\int g f(g,\lambda)\,\ud g=
      \ds\sum_{k=0}^\infty P_\lambda(k)\ds\int g f_k(g)\,\ud g=\\
      \\
      =&\ds\sum_{k=0}^\infty P_\lambda(k)\cdot k \langle G_1\rangle=
      \lambda\langle G_1\rangle.
  \end{array}
$$

In fact, to allow for easy comparison with the data, we consider
the total conductance \emph{per unit area} $\sigma$. By making the
substitution $\sigma=G/A$. It follows directly that the average
value of $\sigma$ satisfies
\begin{equation}
  \langle\sigma\rangle=tN_\ud\langle G_1\rangle.
  \label{eq:avgsig}
\end{equation}
This is a very intuitive result, since $\langle\sigma\rangle$
equals the average number of dopants in the barrier per unit area
multiplied by the average conductance per channel.

To obtain the standard deviation we first observe that
$$
  \begin{array}{rl}
    \langle G_k^2\rangle=&
    \langle(G_1^{(1)}+\ldots+G_1^{(k)})^2\rangle=\\
    \\
    =& k\langle G_1^2\rangle+k(k-1)\langle G_1\rangle^2,
  \end{array}
$$
when the $G_1^{(i)}$ are independently identically distributed.
Using this fact, we find that
$$
  \begin{array}{rl}
    \langle G^2\rangle=&\ds\int g^2 f(g,\lambda)\,\ud g=
      \ds\sum_{k=0}^\infty P_\lambda(k)\ds\int g^2 f_k(g)\,\ud g=\\
      \\
      =&\ds\sum_{k=0}^\infty P_\lambda(k)\cdot
           (k \langle G_1^2\rangle + k(k-1)\langle G_1\rangle^2)=\\
      \\
      =&\lambda\langle G_1^2\rangle + \lambda^2\langle G_1\rangle^2.
  \end{array}
$$
Finally, this yields
$$
  \sd(G)=\sqrt{\langle G^2\rangle -\langle G\rangle^2}=
    \sqrt{\lambda\langle G_1^2\rangle}.
$$

By making the substitution $\sigma=G/A$ once more, we find that
the standard deviation $\sd(\sigma)$ (which can be interpreted as
the spread in $\sigma$) is given by
\begin{equation}
  \sd(\sigma)=\sqrt{\frac{tN_\ud}{A}\langle G_1^2\rangle}.
  \label{eq:sdsig}
\end{equation}
The most important observation from this equation is that
$\sd(\sigma)$ is proportional to $1/\sqrt{A}$, showing that the
area-dependence of $\sd(N)/A$ leads to a similar behavior of
spread in $\sigma$. This also justifies the choice of the
fit-function in Fig.~\ref{fig:meanstd}.

\section{Application to the data}

Our simple model captures the general features of the data and
yields reasonable values for the parameters. To demonstrate this,
a least square fit of the function $C/\sqrt{A}$ (cf.\
Eq.~(\ref{eq:sdsig}); $C$ is a fit-parameter) to the standard
deviations has been performed. The result is the dashed line in
Fig.~\ref{fig:meanstd}. The fit gives a good description of the
data, showing that the spread in the data is consistent with the
prediction of the model.

From the fit, we make some estimates for the physical quantities
in the model. The value of the fit-parameter
$C=1.8$~nA~V$^{-1}$~nm$^{-1}$ should be equal to
$\sqrt{tN_\ud\langle G_1^2\rangle}$ (according to
Eq.~(\ref{eq:sdsig})). By looking at the large area values in
Fig.~\ref{fig:scatterhigh}, we find that $tN_\ud\langle
G_1\rangle\approx 0.04$~nA~V$^{-1}$~nm$^{-2}$
(Eq.~(\ref{eq:avgsig})). Combining these numbers and assuming that
$\langle G_1\rangle^2\approx \langle G_1^2\rangle$ we find values
for parameters in the model. First, $1/tN_\ud \approx
1/(2500~\mathrm{nm}^2)$, which corresponds to an average of one
dopant per 2500~nm$^2$ device area (indicated in
Fig.~\ref{fig:meanstd}). This number is consistent with e.g.\ an
average doping concentration at the interface \footnote{The doping
concentration close to the silicon surface can be more than an
order of magnitude lower than in the bulk due to out-diffusion
during the sample preparation, as confirmed by simulations.}
around $N_\ud=10^{17}$~cm$^{-3}$ and an effective barrier
thickness of $t=2.5$~nm. Note that because $t$ is the thickness of
the barrier region where dopants influence the barrier height, it
is thinner than the total Schottky barrier thickness (a few tens
of nanometers in this case). Second, we find $\langle G_1
\rangle\approx 100$~nA/V for the average conductance per channel.
In order to achieve this value, it is necessary to have e.g.~a
small patch with a local barrier height of approximately $0.15$~eV
and an area \footnote{This estimate was made using the thermionic
emission model [S.~M.~Sze, \emph{Physics of Semiconductor Devices}
(Wiley, New York, 1981), 2nd ed.].} of approximately 10~nm$^2$.
These numbers are consistent with the actual sample parameters.

On average, the conductance per unit area increases for decreasing
island area, as can be seen from Fig.~\ref{fig:scatterhigh}. This
can be explained by the scaling mechanism described in
Ref.~\onlinecite{smit02b}, which predicts that the barrier
thickness decreases with device size for devices that are smaller
than a few times the Debye length. We note that the increasing
value of $\langle\sigma\rangle$ is not the cause of the increment
of $\sd(\sigma)$, since similar measurements on low-doped samples
do not exhibit an increased scatter (not shown).

As demonstrated by our measurements, the discreteness of doping is
easily observed in highly doped samples. In lower doped samples,
where the barrier is thicker, the effect is expected to be much
weaker. The local distortion of the potential landscape due to the
presence of a dopant is roughly as large as its effective
Bohr-radius, which equals about 3~nm in silicon. When the barrier
thickness is much larger than the dopant's potential well, the
effective barrier lowering will be negligible. Hence the induced
local barrier lowering is the most pronounced in thin barriers.

\section{Conclusions}

In conclusion, we have shown that individual dopant atoms dominate
the transport characteristics of epitaxial nanometer sized metal
semiconductor diodes. Room temperature data show increasingly
large device-to-device conductance fluctuations towards smaller
device sizes. Measurements at 4.5~K reveal pronounced structure in
the $IV$-curves. A statistical analysis based on the assumption of
randomly positioned individual dopant atoms leads to a good
description of the experimental data.

\begin{acknowledgments}
We thank S.~G.~Lemay for making the low-temperature measurements
possible. One of us, S.R., acknowledges the Royal Netherlands
Academy of Arts and Sciences for financial support. This work is
part of the research program of the Stichting voor Fundamenteel
Onderzoek der Materie, which is financially supported by the
Nederlandse Organisatie voor Wetenschappelijk Onderzoek.
\end{acknowledgments}

\end{document}